\documentclass[prb,aps,twocolumn,floatfix,superscriptaddress]{revtex4}

\usepackage{graphicx}% Include figure files
\usepackage{dcolumn}% Align table columns on decimal point
\usepackage{bm}% bold math
\usepackage{amssymb}
\usepackage{epsfig}
\usepackage{graphicx}
\usepackage[centertags]{amsmath}

\begin{document}

\title {Contact resistance in graphene-based devices}
\author {S. Russo\footnote{s.russo@tnw.tudelft.nl}}
 \affiliation{Department of Applied Physics, The University of
Tokyo, Tokyo 113-8656, Japan}\affiliation {Kavli Institute of
Nanoscience, Delft University of Technology, Lorentzweg 1, 2628 CJ
Delft, The Netherlands}
\author {M.F. Craciun}
\affiliation{Department of Applied Physics, The University of Tokyo,
Tokyo 113-8656, Japan}
\author{M. Yamamoto}
\affiliation{Department of Applied Physics, The University of Tokyo,
Tokyo 113-8656, Japan}
\author {A. F. Morpurgo}
\affiliation{DPMC and GAP, University of Geneva, quai
Ernest-Ansermet 24, CH-1211 Geneva 4, Switzerland}
\author {S. Tarucha}
\affiliation{Department of Applied Physics, The University of Tokyo,
Tokyo 113-8656, Japan} \affiliation{Quantum Spin Information
Project, ICORP, Japan Science and Technology Agency, Atsugi-shi,
243-0198, Japan}

\begin{abstract}

We report a systematic study of the contact resistance present at
the interface between a metal (Ti) and graphene layers of different,
known thickness. By comparing devices fabricated on 11 graphene
flakes we demonstrate that the contact resistance is quantitatively
the same for single-, bi-, and tri-layer graphene ($\sim800 \pm 200
\Omega \mu m$), and is in all cases independent of gate voltage and
temperature. We argue that the observed behavior is due to charge
transfer from the metal, causing the Fermi level in the graphene
region under the contacts to shift far away from the charge
neutrality point.

\end{abstract}

\maketitle

\newpage

The versatility of graphene-based materials is illustrated by the
large variety of novel electronic phenomena that have been recently
discovered in these systems. Examples are provided by Klein
tunneling in single layers and the opening of a gate tunable band
gap in
bilayers\cite{FLG1,graphene3,Kleinparadox,bilayer1,bilayer2,bilayer3,bilayer4}.
This versatility, together with the surprisingly high values of
carrier mobility\cite{Bolotin} -which exceed by far those of
technologically relevant semiconductors such as Silicon- make
graphene-based materials promising candidates for possible
electronic device applications\cite{graphene3}.\\
Whereas considerable work has focused on the electronic properties
of bulk graphene, virtually no experiments have addressed the
properties of metal/graphene
interfaces\cite{Blake,Lee,Huard,Giovannetti,Malola}. This is
somewhat surprising, since these interfaces will unavoidably be
present in future electronic device, and may crucially affect their
performance. In recently demonstrated single-molecule sensors, for
instance, graphene trilayers have been claimed to be better suited
than single-layers because of a lower contact resistance, leading to
a higher device sensitivity (the measurements of the values of
contact resistance, however, were not discussed in any detail -see
Ref. [4] and related online supporting material). Not only in the
realm of electronic applications, but also for many transport
experiments of fundamental interest, the quality of graphene/metal
contacts is of crucial importance. For example, the simplest
shot-noise measurements require the use of a two terminal
configuration, and it was recently argued\cite{Cayssol} that
properly taking into account the quality of the contacts is
essential to interpret the experimental data correctly.\\
In order to better understand the influence of the contacts we have
performed a series of measurements of the contact resistance
($R_{C}$) present at the interface between Ti/Au electrodes and
graphene layers of different thickness (single, double and triple
layer). The Ti/Au bilayer was chosen because, together with Cr/Au,
it is most commonly used as electrode. In addition, in contrast to
the Cr interlayer, Ti/graphene interface gives highly transmissive
contacts, as demonstrated by the large probability for Andreev
reflection reproducibly observed in Josephson junctions with
Ti/Al\cite{supercurrent}.\\
Our work is based on transport measurements performed on graphene
flakes of different thickness (11 in total: three single layers, six
bilayers, and two trilayers), on which different kinds of devices
were fabricated. Using these devices we succeeded in extracting the
value of contact resistance as a function of gate voltage, using
three different methods: through scaling as a function of device
length, of device width, and by comparing the resistance values
measured in a two and four terminal device configuration. We find
that, irrespective of the method used to extract the contact
resistance, $R_{C} \approx 800 \Omega \mu$m, {\it independent of
thickness of the graphene layer, gate voltage, and temperature}.\\
The graphene flakes utilized in the device fabrication were obtained
by mechanical exfoliation of natural graphite, and subsequently
transferred onto an highly doped $Si$ substrate (acting as a back
gate), coated with a $285$nm $SiO_{2}$ layer. Metallic contacts were
defined by conventional electron- beam lithography, electron-gun
evaporation of Ti/Au ($10/25$nm thick), and lift-off. The thickness
of the graphene layers was identified by determining the shift in
intensity in the RGB green channel relative to the substrate
\cite{graphenevisibility1,graphenevisibility2,graphenevisibility3,bilayer4},
analyzing images taken with a digital camera under an optical
microscope. For a number of flakes the thickness determination was
also confirmed by means of transport measurements (quantum Hall
effect, resistance dependence on a perpendicular electric field,
etc.). Different contact configurations were employed, with two and
four contacts, to enable the quantitative determination of the
contact resistance by both scaling experiments and multi-terminal
measurements. To this end, conducting channel with width $W$ ranging
from $0.8 \mu m$ to $3.5 \mu m$, and contact separation $L$ ranging
from $1.2 \mu m$ to $8.8 \mu m$ were fabricated. The use and the
comparison of these different configurations was instrumental to
insure the uniformity of the graphene layers and of the current
injected from the contact, which are both essential for a reliable
quantitative determination of the contact resistance. All
measurements were taken using a lock-in technique (excitation
frequency: $19.3$Hz), in the linear transport regime, at
temperatures ranging from 50 mK to 300 K, depending on the specific
device.\\

\begin{figure}[h]
\begin{center}
\includegraphics[width=1\columnwidth]{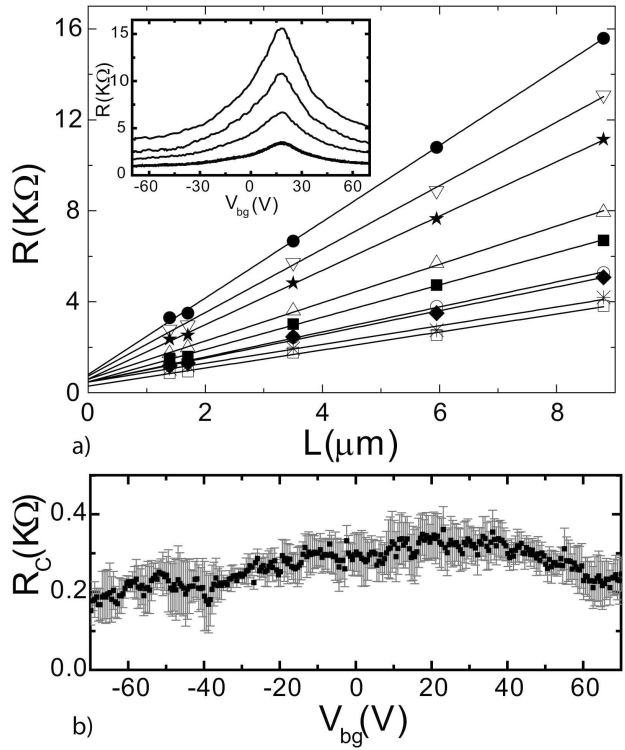}
\end{center}
\noindent{\caption{a) The inset shows the total device resistance
$R_{2p}$ measured at $T=250mK$ in a bilayer graphene with Ti/Au
contacts for fixed device width ($W=5.5 \mu$m) and several different
contact separations (from high to low resistance $L= 1.4, 1.7, 3.5,
6$ and $8.8 \mu$m, respectively). The main panel shows the scaling
of the device resistance \textit{vs.} $L$, for different values of
fixed $V_{bg}$ ($\Box -70V, \ast -50V, \circ -30V, \vartriangle
-10V, \triangledown 10V, \bullet 18V, \star 30V, \blacksquare 50V,
\blacklozenge 70V$). Continuous lines are linear fits to the
experimental data. The contact resistance extracted from the
intercept at L=0 at each fixed $V_{bg}$ in the range
$-70V<V_{bg}<70V$ is shown in panel (c).} \label{Fig1}}
\end{figure}

\begin{figure}[h]
\begin{center}
\includegraphics[width=1\columnwidth]{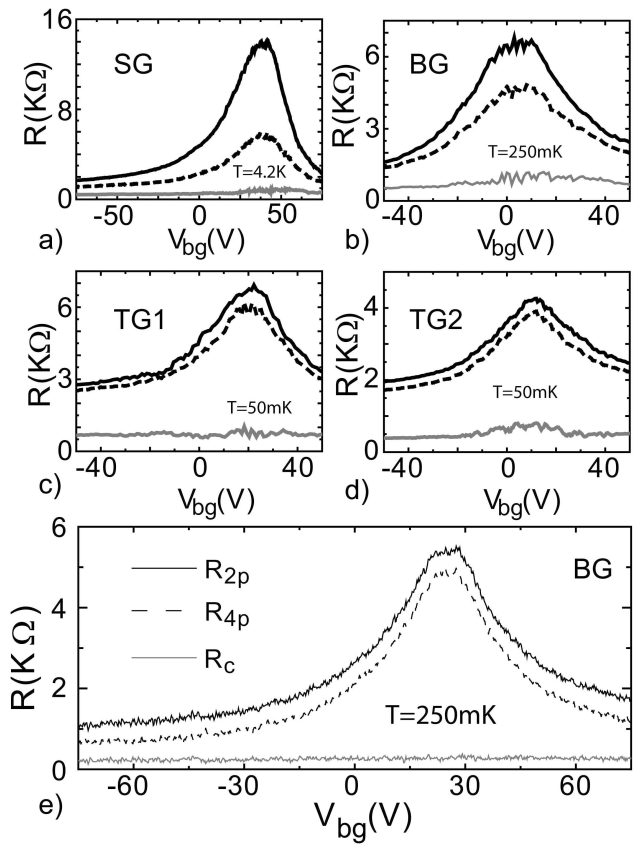}
\end{center}
\noindent{\caption{Gate-voltage dependence of $R_{C}$ (light gray
curve) extracted from the scaling with device width of $R_{2p}$, on
single layers in (a) (with $L=2.75 \mu m$, $W=0.8 $ and $2.4 \mu m$
respectively for the continuous and dashed line measurements), on
double layers in (b) ($L=1.26 \mu m$, $W=1.05$ and $1.8 \mu m$
respectively for the continuous and dashed line measurements), and
on trilayers in (c,d) (in (c) $L=1.2 \mu m$, $W=1.62$ and $1.94 \mu
m$ respectively for the continuous and dashed line, in (d) $L=1.25
\mu m$, $W=1.66$ and $2.12 \mu m$ respectively for the continuous
and dashed line measurement).e) Gate-voltage dependence of $R_{C}$
obtained from the comparison of two and four probe measurements, as
described in the text, for a double layer device ($W=3.3\mu m L=1.96
\mu m$).} \label{Fig2}}
\end{figure}

One of the methods most commonly used to determine the contribution
of the resistance present at an interface between two different
materials is by means of a scaling analysis of the resistance,
measured in a two probe configuration in devices with different
contact separation. Specifically, the two-probe resistance of a
graphene device reads $R_{2p}(Vbg) = 2R_{C}(V_{bg}) +
R_{G}(V_{bg})$, where $R_{G} = \rho_{G}(V_{bg})L/W$ is the
contribution of graphene to the resistance ($\rho_{G}(V_{bg})$
graphene resistivity) and $ R_{C}(V_{bg})$ is the (contact)
resistance of one metal/graphene interface. Experimentally,
$R_{C}(V_{bg})$ is obtained by measuring the resistance of devices
having different lengths $L$, and extrapolating the data to $L=0$
(while keeping fixed W).\\
The inset of Fig. \ref{Fig1}a shows measurements of $R_{2p}(V_{bg})$
performed on devices fabricated on a bilayer graphene flake, with
different contacts separations ($L$ ranging from $1.4 \mu m$ to $8.8
\mu m$) and fixed conductive channel width ($W=5.5 \mu m$). As it
appears from the main graph in Fig. \ref{Fig1}a, at each fixed value
of $V_{bg}$ the total device resistance scales linearly with $L$.
The deviations from such a linear dependence are small, indicating
that the contact resistance for the different electrodes is
approximately the same. From the linear extrapolation of $R_{2p}$ we
determine the intercept at $L=0$ as a function of $V_{bg}$. It
appears that $R_{C}$ is only weakly dependent on $V_{bg}$ even in
the charge neutrality region (see Fig. \ref{Fig1}b), in contrast to
the resistance of bilayer graphene, which exhibits a pronounced
peak.\\
We have also checked the scaling as a function of contact width but
fixed channel length, by comparing two devices fabricated on the
same flake. In this case $R_{C}$ is given by
$(R_{2p}^{Dev1}-\rho_{G}(V_{bg})L^{Dev1}/W^{Dev1})/2$, with
$\rho_{G}(V_{bg})=(R_{2p}^{Dev1}-R_{2p}^{Dev2})(L^{Dev1}/W^{Dev1}-
L^{Dev2}/W^{Dev2})^{-1}$. In Fig. \ref{Fig2}a-d we show the results
of this experiments for layers of different thickness, with the
light grey lines representing values obtained for $R_{C}$ as a
function of $V_{bg}$. Consistently with the previous results, also
these experiments show that $R_{C}$ is a gate independent quantity
over the full back gate range ($\sim -70V<V_{bg}<70V$), and that its
value ($\sim 800 \Omega \mu m$) does not depend on the thickness of
graphene layer.\\

Finally, we have extracted the value of $R_{C}$ by comparing
directly two and four probe resistance measurements. In a four-probe
configuration only the resistance of the graphene channel is
measured, i.e. $R_{G}(V_{bg})=R_{4p}$. From the value of $R_{4p}$
and the known device geometry we obtain the resistivity of graphene,
and use it to extract the contact resistance from resistance
measured in a two-terminal configuration $R_{2p}$. In Fig.
\ref{Fig2}e we plot $R_{2p}$ and $R_{4p}$ versus $V_{bg}$, together
with the extracted $R_{C}$. Once again we find that $R_{C} \sim 800
\Omega \mu m$ and gate voltage independent. The fact that all these
three independent transport methods (scaling of $L$, $W$, and
comparison of two- and four-probe measurements) give quantitatively
consistent results confirms the validity of our analysis. Note also
that measurements performed at different temperature give the same
result, indicating that contact resistance is temperature
independent (or only very weakly temperature dependent) up to room
temperature.\\

\begin{figure}[h]
\begin{center}
\includegraphics[width=1\columnwidth]{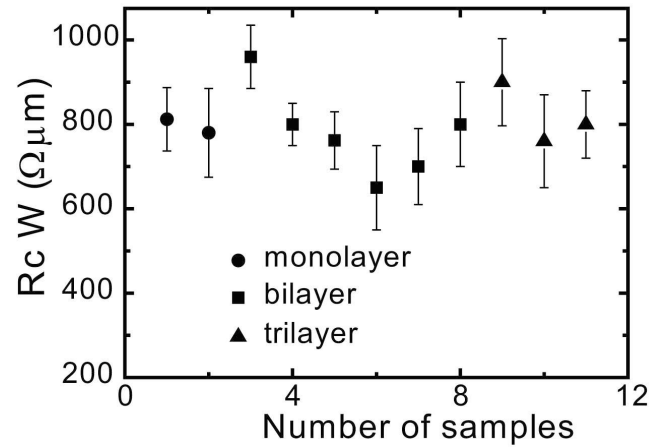}
\end{center}
\noindent{\caption{Summary plot of $R_{C}$ estimated for 11
different few layer graphene devices.} \label{Fig3}}
\end{figure}

A remarkable result of our measurements emerges when comparing the
estimated value of $R_{C}$ for each different few layer graphene
device (see Fig. \ref{Fig3}). Even though graphene-based materials
of different thickness correspond to truly different electronic
systems, with unique and characteristic low-energy electronic
properties, the value of $R_{C}$ that we have obtained from all our
measurements is independent of the number of layers: at least up to
3 layers $R_{C} = 800 \pm 200 \Omega \mu m$. Since the low-energy
electronic properties of single-, bi-, and tri-layer are markedly
different\cite{Latil}, the independence of $R_{C}$ from layer
thickness suggests that a substantial charge transfer from the metal
contact to the graphene shifts the Fermi level far from the
degeneracy point. This same argument may also explain why $R_{C}$ is
independent on $V_{bg}$, since the density of charge transferred
from the metal contact can easily be much larger than the typical
modulation induced by the back gate voltage. Indeed, it has been
predicted theoretically that a large transfer of charge should occur
between many different metals and graphene\cite{Giovannetti}. For
Ti, however, no calculations have been yet performed.\\
In conclusion we have conducted a systematic study in transport
experiments of the contact resistance at graphene-metal (Ti/Au)
interface, using single, double and triple layer graphene. Employing
three independent methods we have established that $R_{C}$ is $\sim
800 \pm 200 \Omega \mu m$, independent of back gate voltage, of
temperature and of layer thickness. A significant charge transfer at
the graphene-metal interface, which shifts the Fermi level of the
few-layer graphene far away from degeneracy point, is the likely
explanation for this unexpected result.\\
We acknowledge financial support from FOM (AFM and SR) and the Japan
Society for the Promotion of Science, grant P07372 (MFC). M.Y.
acknowledge financial support from the Grant-in-Aid for Young
Scientists A (No. 20684011) and ERATO-JST (No. 080300000477). S.T.
acknowledges financial support from the Grant-in-Aid for Scientific
Research S (No.19104007), B (No. 18340081), JST-CREST, and Special
Coordination Funds for Promoting Science and Technology.\\


\begin{thebibliography}{10}

\bibitem{FLG1} K. S. Novoselov, A. K. Geim, S. V. Morozov, D. Jiang, Y. Zhang,
S. V. Dubonos, I. V. Grigorieva, and A. A. Firsov, \textit{Science}
\textbf{306}, 666 (2004); K. S. Novoselov, A. K. Geim, S. V.
Morozov, D. Jiang, M. I. Katsnelson, I. V. Grigorieva, S. V.
Dubonos, and A. A. Firsov, \textit{Nature} \textbf{438}, 197 (2005);
Y. Zhang, Y. Tan, H. L. Stormer, and P. Kim, \textit{Nature}
\textbf{438}, 201 (2005).


\bibitem{graphene3} A. K. Geim and K. S. Novoselov, \textit{Nature Mater.}
\textbf{6}, 183 (2007).

\bibitem{Kleinparadox} M. I. Katsnelson, K. S. Novoselov, and A. K. Geim,
\textit{Nature Physics} \textbf{2}, 620 (2006).

\bibitem{bilayer1} E. McCann, \textit{Phys. Rev. B} \textbf{74}, 161403 (2006).

\bibitem{bilayer2} T. Ohta, A. Bostwick, T. Seyller, K. Horn, and E. Rotenberg,
\textit{Science} \textbf{313}, 951 (2006).

\bibitem{bilayer3} E. V. Castro, K. S. Novoselov, S. V. Morozov, N. M. R. Peres,
J. M. B. Lopes dos Santos, Johan Nilsson, F. Guinea, A. K. Geim, and
A. H. Castro Neto, \textit{Phys. Rev. Lett.} \textbf{99} 216802
(2007).

\bibitem{bilayer4} J. B. Oostinga, H. B. Heersche, X. Liu, A. F. Morpurgo, and
L. M. K. Vandersypen, \textit{Nature Mater.} \textbf{7}, 151 (2008).

\bibitem{Bolotin} K.I. Bolotin, K.J. Sike, Z. Jian, G. Fundenberg, J. Hone, P. Kim and H.L. Stormer
\textit{Solid State Commun.} \textbf{146}, 351-355 (2008).

\bibitem{Blake} P. Blake, R. Yang, S. V. Morozov, F. Schedin, L. A.
Ponomarenko, A. A. Zhukov, I. V. Grigorieva, K. S. Novoselov, A. K.
Geim, arxiv:0811.1459.

\bibitem{Lee} E. J. H. Lee, K. Balasubramanian, R. T. Weitz,
M. Burghard, and K. Kern, \textit{Nature Nanotechnology} \textbf{3},
486 (2008).

\bibitem{Huard} B. Huard, N. Stander, J. A. Sulpizio, and D.
Goldhaber Gordon, \textit{Phys. Rev. B} \textbf{78}, 121402(R)
(2008).

\bibitem{Giovannetti} G. Giovannetti, P. A. Khomyakov, G. Brocks, V. M.
Karpan, J. van den Brink, and P. J. Kelly, \textit{Phys. Rev. Lett.}
\textbf{101}, 026803 (2008).

\bibitem{Malola} S. Malola, H. Hakkinen, and P. Koskinen,
arxiv:0811.1459.

\bibitem{Cayssol} J. Cayssol, B. Huard, and D.
Goldhaber-Gordon, Preprint at <http://arxiv.org/abs/0810.4568>
(2008).

\bibitem{supercurrent}H.H. Heersche, P. Jarillo-Herrero, J.B.
Oostinga, L.M.K. Vandersypen, and A.F. Morpurgo, \textit{Nature}
\textbf{446}, 56-59 (2007).

\bibitem{graphenevisibility1}  P. Blake, K. S. Novoselov, A. H. Castro Neto, D.
Jiang, R. Yang, T. J. Booth, A. K. Geim, E. W. Hill, \textit{ Appl.
Phys. Lett.} \textbf{91}, 063124 (2007).

\bibitem{graphenevisibility2} D. S. L. Abergel, A. Russell, and V. I. Falko,
\textit{Appl. Phys. Lett.} \textbf{91}, 063125 (2007).

\bibitem{graphenevisibility3} Z. H. Ni, H. M. Wang, J. Kasim, H. M. Fan, T. Yu,
Y. H. Wu, Y. P. Feng, and Z. X. Shen, \textit{Nano Letters}
\textbf{7}, 2758-2763 (2007).

\bibitem{contactresistence5} R. Golizadeh-Mojarad, and S. Datta, arxiv:0710.2727.

\bibitem{Latil} S. Latil, L. Henrard, \textit{Phys. Rev. Lett.} \textbf{97}, 036803
(2006); B. Partoens, F.M. Peeters, \textit{Phys. Rev. B}
\textbf{74}, 075404 (2006).

\end{thebibliography}
\end{document}